\documentclass[%
 reprint,
 amsmath,amssymb,
 aps,
]{revtex4-2}

\usepackage{graphicx}
\usepackage{dcolumn}
\usepackage{bm}

\begin{document}

\preprint{APS/123-QED}

\title{Two-cavity-mediated photon-pair emission by one atom}

\author{Gianvito Chiarella$^{1,2}$}
\author{Tobias Frank$^{1,2}$}
\author{Pau Farrera$^{1,2}$}
    \email{pau.farrera@mpq.mpg.de}
    
\author{Gerhard Rempe$^{1,2}$}
\affiliation{$^{1}$Max-Planck-Institut für Quantenoptik, Hans-Kopfermann-Straße 1, 85748 Garching, Germany}
\affiliation{$^{2}$Munich Center for Quantum Science and Technology (MCQST), Schellingstr. 4, D-80799 München}




\begin{abstract}
Photon-pair sources are widely used in quantum optics and quantum information experiments. Despite their broad deployment, there has not yet been an on-demand implementation with efficient into-fiber photon generation and high single-photon purity. Here we report on such a source based on a single atom with three energy levels in ladder configuration and coupled to two optical fiber cavities. We efficiently generate photon pairs with in-fiber emission efficiency of $\eta_{\mathrm{pair}}=16(1)\%$ and study their temporal correlation properties. We simulate theoretically a regime with strong atom-cavity coupling and find that photons are directly emitted from the ground state, i.e. without atomic population in any intermediate state. We propose a scenario to observe such a double-vacuum-stimulated effect experimentally.
\end{abstract}

\maketitle

\noindent The development of light sources that generate pairs of single photons has played a major role in quantum science and technology. On the fundamental side they have been used for basic tests in quantum optics and quantum information such as early demonstrations of Bell inequality violation \cite{Clauser1978, Aspect1981}. On the applied side they have been used as heralded single photon sources \cite{Hong1986, Grangier1987, Rarity1987} and are currently used in emerging applications such as quantum cryptography \cite{Gisin2012}, photonic quantum computing \cite{Zhong2020}, or quantum metrology \cite{Walther2004}. In any of these applications it is important to generate the photons efficiently and to have photon number states consisting on pairs of pure single photons. Especially for applications in quantum communication protocols, the photons need to be efficiently collected by and routed through optical fibers. Additionally, protocols including the interference between photons generated by different network nodes require high single-photon indistinguishability \cite{Zeilinger1997, Duan2001}.

The broad range of possibilities has motivated the development of a variety of implementations based on different physical systems. Initial experiments were performed using cascaded photon emission in atomic vapours \cite{Kocher1967}, and later spontaneous parametric down conversion (SPDC) in nonlinear crystals \cite{Burnham1970, Zhang2021}. Sources of narrow-band photon pairs have also been implemented using four-wave mixing in cold atomic clouds \cite{Srivathsan2013}. However, the nature of the photonic state generated by all these approaches provides a fundamental trade-off between the photon emission efficiency and the generation of multiphoton components. Further approaches make use of cascaded emission in single molecules \cite{Rezai2019} and in quantum dots \cite{Stevenson2006, Huber2018}. More recently, biexciton-exciton cascaded processes in quantum dots embedded in optical cavities have reached high collection efficiencies and excellent photon number purity \cite{Dousse2010, Wang2019}. This system is in principle a good candidate for an efficient generation of pairs of pure single photons. However, so far both photons are emitted into different modes of the same optical resonator. This poses limitations in the efficient coupling of both fields into a single mode fiber \cite{Dousse2010} and on future prospects for obtaining high photon indistinguishability \cite{Scholl2020,Sbresny2022}. As we explain below, photon indistiguishability requires independent control of the photon emission in each transition, which is hard to achieve with a single cavity. 
Simultaneously accomplishing all the aforementioned features makes the realization of the wanted photon-pair source highly challenging, and despite numerous efforts, its implementation remains unaccomplished.

\begin{figure}[b!]
\centering
\includegraphics[width=1\linewidth]{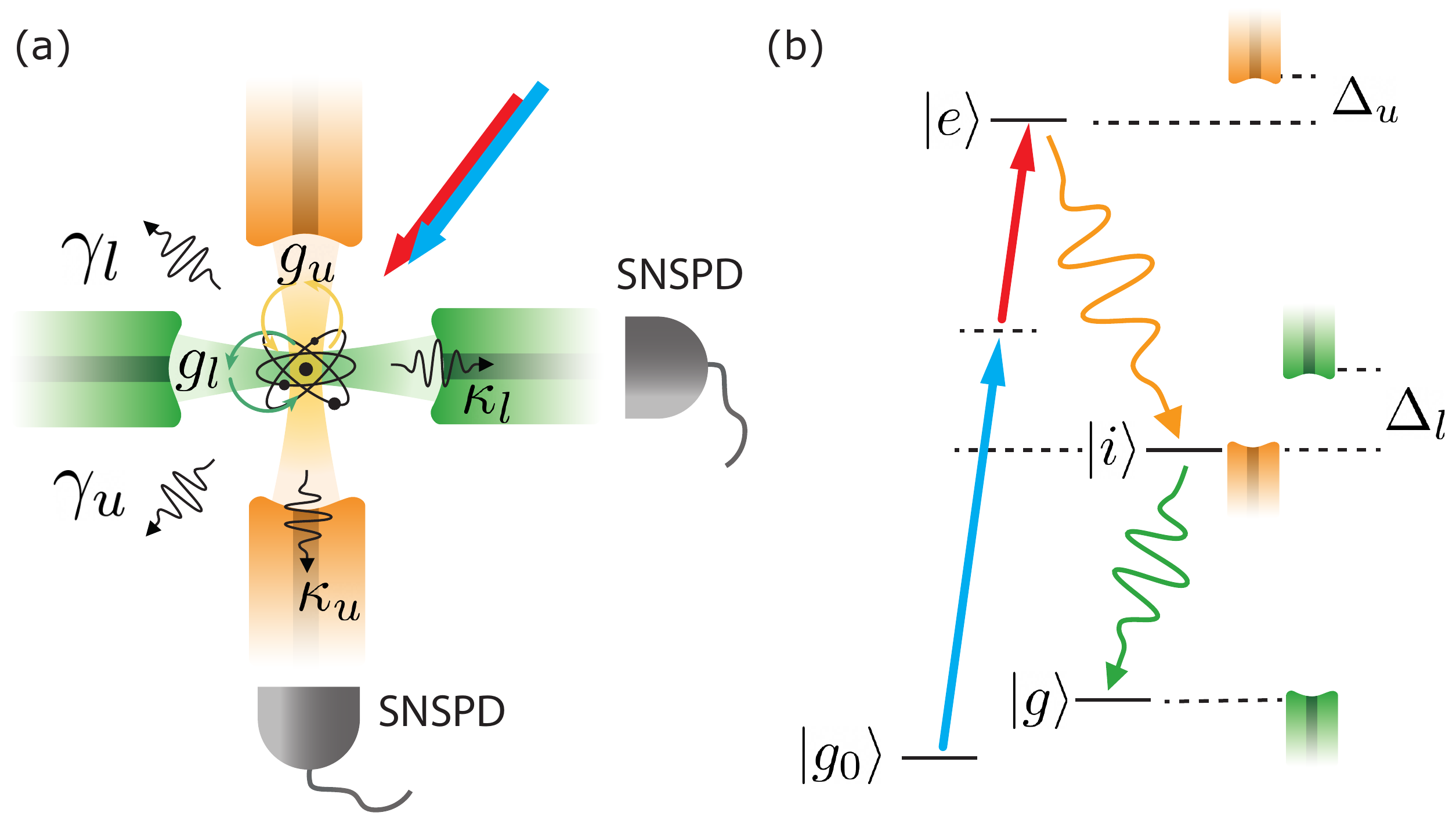}
\caption{(a) A single atom trapped at the center of two crossed fiber cavities. The cavities are single-sided so that photons escape the cavity predominantly through one of the mirrors. (b) The two cavities are coupled to two cascaded transitions in a ladder configuration ($\left|g\right>-\left|i\right>-\left|e\right>$). An additional ground state $\left|g_0\right>$ is used in order to initialize and excite the atom to state $\left|e\right>$ .}
\label{fig:intro}
\end{figure}

Here we report on a new source of photon pairs based on a single neutral atom coupled to two independent optical cavities (Fig.~\ref{fig:intro}a). The principle advantage of the approach is that the two photons can individually be controlled by the parameters of the two cavities. The atom has three energy levels in ladder configuration, supporting two transitions that are coupled to two optical fiber cavities in the regime of high cooperativity (Fig.~\ref{fig:intro}b). A drive laser couples the ground and excited states, so that two single photons can be emitted, one in each of the two cavity fibers. We show that the two photons are temporally correlated, and propose a set of atom-cavity parameters for which the photons coming from the same transition of two identical sources are indistinguishable. We also simulate theoretically the regime in which the cavities are coupled to the respective transitions in the strong coupling regime (where the atom-cavity coupling rate $g$ is larger than the spontaneous dipole emission rate $\gamma$ and the cavity-field decay rate $\kappa$). We observe that in this regime a transition to the ground state can happen without populating any intermediate state, in a process similar to Stimulated Raman Adiabatic Passage (STIRAP) \cite{Bergmann1998}, but mediated by two cavity-vacuum fields. Finally we propose a set of parameters in order to observe this effect in an experiment. 

\begin{figure}[t]
\centering
\includegraphics[width=1\linewidth]{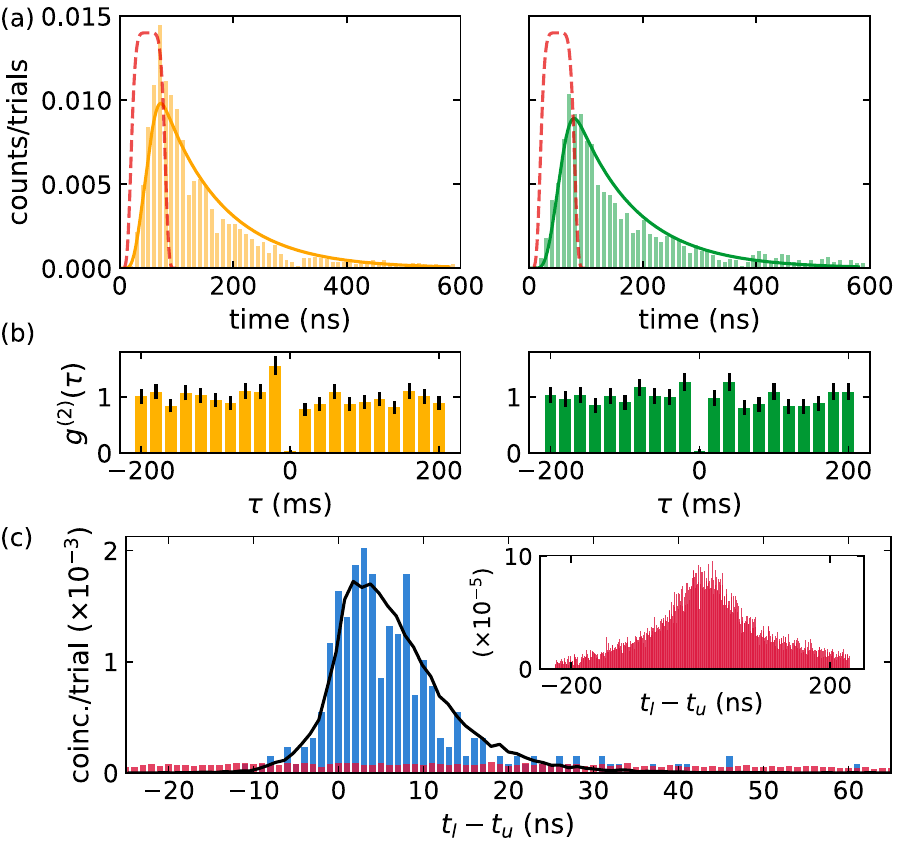}
\caption{(a) Time histograms showing photon counts detected at the outputs of the upper (orange) and lower (green) cavities after exciting the atom to state $\left|e\right>$ at time zero. The solid lines show the photons' temporal profile predicted by the theoretical simulation. The red dashed line represents the profile of the excitation pulses. (b) Auto-correlation function for the photons emitted from the upper (orange) and the lower (green) cavities. (c) Cross-correlation as a function of the delay between two photon detection events. The solid black line indicates the curve predicted by the theoretical model. The red bars in both the main figure and the inset show the coincidence time histogram between photons emitted in different trials. For this data the correlation peak has been arbitrarily centered to $t_l-t_u=0$.}
\label{fig:results}
\end{figure}

In our experiment we optically trap single $\mathrm{^{87}Rb}$ atoms at the center of two crossed optical fiber cavities \cite{Brekenfeld2020}. The three-level system that we use to generate the photon pairs consists of the ground state $\left|g\right>=\left|5^2S_{1/2},F=2\right>$, intermediate state $\left|i\right>=\left|5^2P_{3/2},F=2\right>$ and excited state $\left|e\right>=\left|5^2D_{5/2}, F=3\right>$. One cavity is resonant to the $\left|g\right>-\left|i\right>$ transition at 780~nm, while the other one is coupled to the $\left|i\right>-\left|e\right>$ transition at 776~nm. Both cavities have one mirror with a significantly higher transmission than the other (340~ppm and 10~ppm), such that the intracavity fields escape the cavities dominantly through the more transparent mirrors. After trapping and cooling single atoms we initialize them in state $\left|g_0\right>=\left|5^2S_{1/2},F=1\right>$, we then apply two pulses with duration of 60 ns that couple the transitions $\left|g_0\right>-\left|i\right>$ and $\left|i\right>-\left|e\right>$, and excite the atom in $\left|e\right>$. During this time we switch off the optical trap, in order to avoid photo-ionization of the atom (see Appendix B).

\begin{figure}[t]
\centering
\includegraphics[width=1\linewidth]{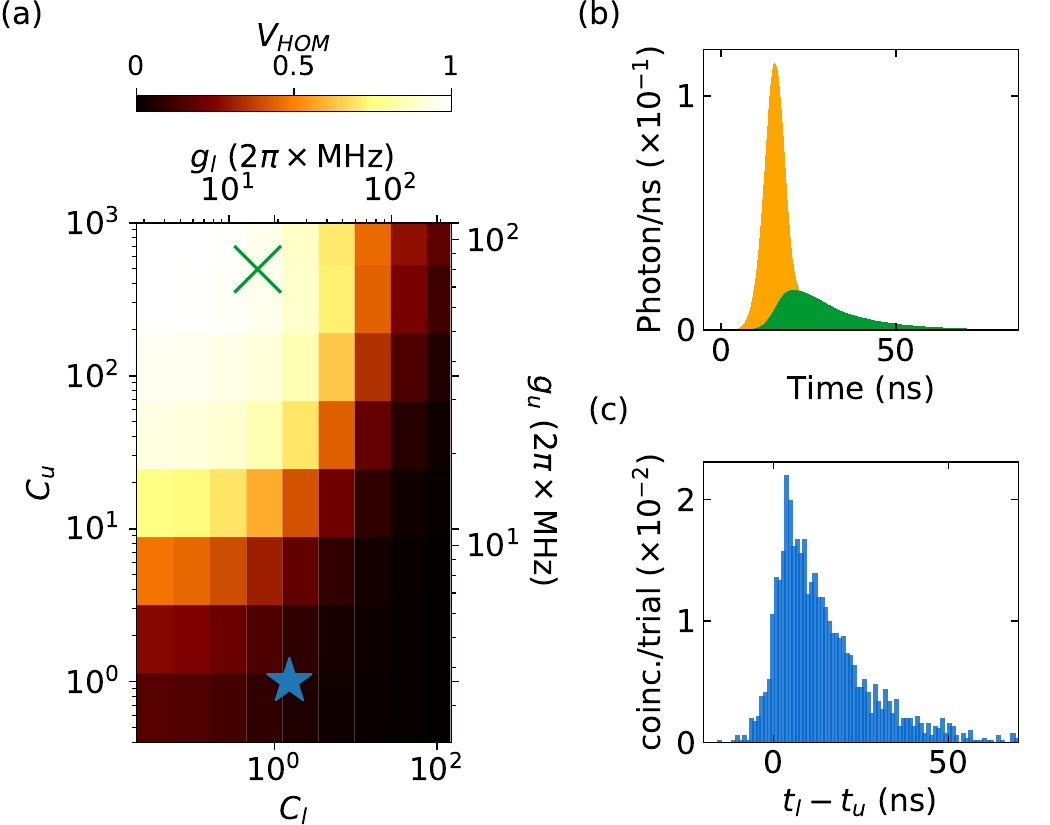}
\caption{(a) Simulated Hong-Ou-Mandel interference visibility ($V_{\mathrm{HOM}}$) for the upper photons as a function of the atom-cavity coupling rate of the upper ($g_u$) and lower ($g_l$) cavity. The atom-cavity cooperativities $C_u$ and $C_l$ corresponding to each coupling rate $g_u$ and $g_l$ are also indicated. The blue star marks our experimental parameters $(g_u, g_l)$ = (4, 21.9) $\times 2\pi \mathrm{MHz}$, for which  $V_{\mathrm{HOM}}=0.08$. Plot (b) and (c) show, respectively, the simulated intensity profiles and cross-correlation for the cavity parameters marked by the green cross in the main plot $(g_u, g_l)$ = (80, 15) $\times 2\pi \mathrm{MHz}$. For these parameters $V_{\mathrm{HOM}}=0.95$.}
\label{fig:HOM}
\end{figure}

Fig.~\ref{fig:results}a shows the photon-count versus time histogram corresponding to both the upper and lower cavity fields, after sending the two pulses that drive the transition $\left|g_0\right>-\left|e\right>$. Since the lifetime of state $\left|e\right>$ is much longer than the lifetime of state $\left|i\right>$ the duration of both photons is given by the Purcell shortened lifetime of state $\left|e\right>$ (in free space $\tau_{\left|e\right>}=231(8)~\mathrm{ns}$ \cite{Safronova2011}). The photon emission exhibits lifetimes of  $\tau_u=102(3)~\mathrm{ns}$ and $\tau_l=109(2)~\mathrm{ns}$. By simulating the system with the methods explained in the next paragraphs and in Appendix C, we find these values to be in good agreement with those predicted by the theory ($\tau_u=\tau_l=106~\mathrm{ns}$).
The atom-cavities parameters for the upper cavity are ($ g_u, \gamma_u, \kappa_u) = (4, 0.33, 30)\times 2\pi \mathrm{MHz}$ and those for the lower cavity are ($ g_l, \gamma_l, \kappa_l) = (21.9, 3, 60)\times 2\pi \mathrm{MHz}$, where $g_u$ ($g_l$) is the coupling rate of the cavity that couples to the upper $\left|i\right>-\left|e\right>$ (lower $\left|g\right>-\left|i\right>$) transition, $\gamma_u$ ($\gamma_l$) is the spontaneous dipole emission rate of the upper (lower) transition and $\kappa_u$ ($\kappa_l$) is the cavity-field decay rate of the corresponding cavity. From these parameters we derive cooperativities of $C_u=g_u^2/(2\gamma_u\kappa_u) = 0.8$ for the upper transition and $C_l=g_l^2/(2\gamma_l\kappa_l) =1.33$ for the lower one. The Purcell decay time for the excited state derived from our cooperativity is $\tau_{\left|e\right> \mathcal{P}} = \tau_{\left|e\right>}/(2C_u + 1) = 92(5)$ ns. The discrepancy between this value and the measured one $\tau_u = 102(3)$ ns (in good agreement with the numerical simulations) is because the intermediate state is not infinitely long-lived. The decay from the intermediate state into the ground state degrades the coherent dynamics between the excited and intermediate states mediated by the upper cavity, reducing the emission rate and photon emission efficiency in the upper cavity. 

We observe that we generate photon-pairs with an in-fiber efficiency of $\eta_{\mathrm{pair}}=16(1)\%$ and single photons with in-fiber efficiencies of $\eta_u=40(3)\%$ in the upper cavity and $\eta_l=29(2)\%$ in the lower cavity. The generation efficiency of a photon-pair is greater than the product of the single photons efficiencies ($\eta_u \cdot \eta_l = 12(1)\%$), indicating that the photons are correlated.
The probability that a photon is emitted into the upper (lower) cavity fiber once a photon was detected from the lower (upper) cavity is $\eta_{u|l}=56(4)\%$ ($\eta_{l|u}=41(3)\%$). The inefficiencies include the cavity-fiber mode matching ($\eta_{\mathrm{mm}}^u=0.94$ and $\eta_{\mathrm{mm}}^l=0.81$), the cavity outcoupling efficiency ($\eta_{\mathrm{oc}}^u=0.79$ and $\eta_{\mathrm{oc}}^l=0.85$) and the emission into the cavity mode considering the finite cooperativity ($\eta_{C}^u=0.62$ and $\eta_{C}^l=0.73$). The cavity-fiber mode matching is determined by the cavity mirrors' radia of curvature, and the cavity outcoupling efficiency is given by the ratio between the transmission of the outcoupling mirror and the total cavity losses. Future experiments could envision lower intracavity losses and a better choice of these parameters in order to increase the photonic collection efficiency in the fibers. Moreover, the choice of stronger transitions would increase the cooperativity and therefore the emission efficiency into the cavity mode. A more detailed discussion on fiber cavities design limitations and challenges can be found in Appendix E. We observe that the photon statistics of the fields in each cavity show autocorrelation values of $g_u^{(2)}(0)=0.04(3)$ and $g_l^{(2)}(0)=0.03(2)$, indicating the single-photon purity of both fields (Fig.~\ref{fig:results}b). This is limited by the presence of more than one emitter in the resonators. Even though we trap mainly single atoms at the center of the resonators, the probability of trapping more than one atom is not negligible. From the measured $g_u^{(2)}(0)$ and $g_l^{(2)}(0)$ we infer an average of 1.04(3) and 1.03(2) atoms in the upper and lower cavity, respectively. Looking at the photon coincidences we observe a correlation peak with different rise and fall times (Fig.~\ref{fig:results}c). The rise time ($\tau=2.7(2)~\mathrm{ns}$) is determined by the optical lifetime of the upper cavity ($\tau=2.5~\mathrm{ns}$) while the fall time ($\tau=7.6(4)~\mathrm{ns}$) is given by the Purcell-enhanced decay from state $\left|i\right>$ (in free space $\tau_{\left|i\right>}=26.2~\mathrm{ns}$), matching well the expected value $\tau_{\left|i\right> \mathcal{P}} = \tau_{\left|i\right>}/ (2C_l + 1) = 7.1(2)$~ns. The simulation based on our theoretical model predicts $\eta_{u|l} = 0.63(2)$ and $\eta_{l|u} = 0.49(2)$, indicating that it slightly overestimates the experimentally measured efficiencies. This discrepancy can be attributed to several limitations inherent in the theoretical framework. Firstly, the model does not account for the presence of multiple atoms, which can lead to the detection of uncorrelated photons. Secondly, the excitation process in the model is simplified to consist of a single pulse, rather than incorporating the more complex coupling to the intermediate virtual state. Lastly, the model neglects other hyperfine excited states in $5^2D_{5/2}$, which can be populated during the excitation process.

The temporal width of the two-photon cross-correlation peak is significantly shorter than the duration of both upper and lower cavity single-photon wave packets, showing that the two-photon emission is temporally correlated. Since the emission time of one photon tells information about the emission time of the other photon, the Hong-Ou-Mandel (HOM) interference visibility for each single photon is smaller than unity. Indeed, by simulating the photon-pair emission with the methods explained in the next paragraphs, we can estimate a HOM interference visibility of $V_{\mathrm{HOM}}=|g^{(1)}(0)|^2=0.08$ for both cavity fields \cite{Scholl2020}. However, having two independent cavities coupled to the atom allows us in principle to independently tune the atom-cavity coupling parameters for each transition. By doing so a set of parameters can be found for which the upper and lower cavity fields are generated with no temporal correlation. Fig.~\ref{fig:HOM} shows the simulated $V_{\mathrm{HOM}}$ of the upper photon as a function of the coupling rates of the cavities $g_u$ and $g_l$. The other atom-cavity parameters are set to the values used in the experiment and a pulse with a duration of 7~ns is considered. The interference visibility is determined by the temporal width of the two-photon cross-correlation compared to the temporal width of the single photon wave packets. In other words, if the temporal width of the cross-correlation between two photons emitted in the same trial is the same as the one between photons emitted in different trials, the emission of a photon does not tell any information about the other one. Since the emission rate in each cavity is controlled by the respective coupling strengths, the ratio between $g_u$ and $g_l$ is a critical factor. As shown in Fig.~\ref{fig:HOM}(a), for a fixed $g_u$, an increasing $g_l$ corresponds to a decreasing $V_{\mathrm{HOM}}$. This is because the higher $g_l$, the faster the emission in the lower cavity will take place, giving rise to a temporal correlation between the two photons.  By properly tuning the coupling strengths, one can achieve a HOM interference visibility for the two fields close to unity, as indicated by the green cross in Fig.~\ref{fig:HOM}(a). This point corresponds to the cavities couplings that provide high indistinguishability ($V_{\mathrm{HOM}} \geq 95\%$) and photon-pair efficiency comparable to our measurements.  

Being able to independently tune the cavity parameters is useful both for manipulating the properties of the fields generated and for investigating the atom-field dynamics in different regimes. In particular, one can extend the study of this system to the regime of strong atom-cavity coupling (in which $g>(\kappa,\gamma$)). In this situation, the system exhibits dynamics that is analog to STIRAP. Since modifying the cavity parameters to be in the strong-coupling regime goes beyond our present experimental capabilities, we simulate theoretically the regime of interest. Our cavities are designed with a mirror that has a high transmission (340~ppm) leading to a high $\kappa$, however one could in the future reach the strong coupling regime by using shorter cavities with lower transmission mirrors \cite{Hunger2010}, and using a stronger upper transition such as the $5^2P_{3/2}-4^2D_{5/2}$ $^{87}\mathrm{Rb}$ line.

\begin{figure}[t]
\centering
\includegraphics[width=1\linewidth]{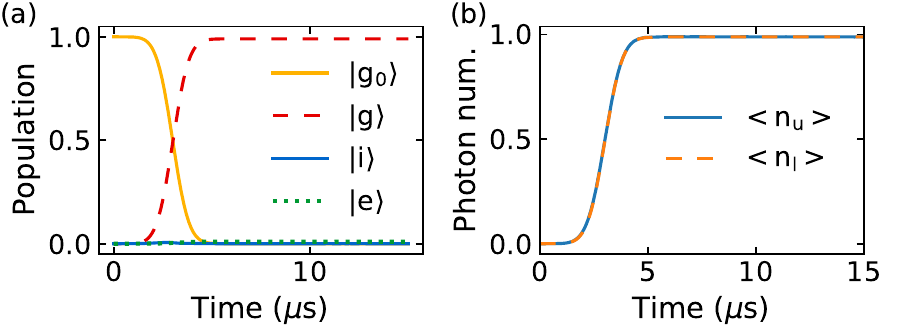}
\caption{Temporal evolution of the atomic populations and the cavity photon numbers. (a) The atom initially in the $\left|g_0\right>$ is coupled to the state of the atom-cavities system $\left|\Psi_0\right>$ by a gaussian pulsed drive with duration 2$\mu s$ for $(g_u, g_l) = (10, 1) \times 2\pi \mathrm{MHz}$. While the atom is quasi-adiabatically transferred from $\left|g_0\right>$ to $\left|g\right>$, a photon is output by each cavity (b). For this scenario, the cavities are resonant to the atomic transitions and atomic and cavity decays rates are small in comparison to the atom-photon coupling rates ($\gamma_u, \gamma_l, \kappa_u, \kappa_l) = (10^{-3}, 10^{-2}, 10^{-10}, 10^{-10})\times 2\pi \mathrm{MHz}$).}
\label{fig:STIRAP}
\end{figure}

We start by considering the Jaynes-Cummings Hamiltonian describing the three atomic levels and the two photonic cavity modes interacting with the upper ($u$) and lower ($l$) transition \cite{Yoo1985}. In the interaction picture the Hamiltonian can be written

\begin{equation}
H_0 = \Bigl(g_l a_l^{\dagger} \sigma_{gi} + g_u a_u^{\dagger} \sigma_{ie} + \mathrm{H.c.}\Bigr) + \Delta_l a_l^{\dagger} a_l + \Delta_u a_u^{\dagger} a_u
\label{eq:hamiltonian}
\end{equation}

\noindent where $\hbar = 1$ and $\Delta_l$ ($\Delta_u$) is the cavity detuning to the lower (upper) atomic transition. Here $a_u$ ($a_l$) and $a^\dag_u$ ($a^\dag_l$) are the annihilation and creation operators of the upper (lower) cavity mode, $\sigma_{gi}$ and $\sigma_{ie}$ are the atomic operators which describe the transition from $\left|i\right>$ to $\left|g\right>$ and from $\left|e\right>$ to $\left|i\right>$, respectively, and H.c. stands for the Hermitian conjugate.

For $\Delta_l=-\Delta_u = \Delta$ the Hamiltonian has three eigenstates with eigenenergies $E_0=0$, $E_1=\frac{\Delta}{2}+\sqrt{g_u^2+g_l^2+(\frac{\Delta}{2})^2}$ and $E_2=\frac{\Delta}{2}-\sqrt{g_u^2+g_l^2+(\frac{\Delta}{2})^2}$, referenced to the energy of state $\left|e\right>$. The eigenstate corresponding to eigenenergy $E_0$ is

\begin{equation}
\left|\Psi_0\right>=\Bigl(-g_l\left|e,0_u,0_l\right>+g_u\left|g,1_u,1_l\right>\Bigr)/\sqrt{g_l^2+ g_u^2}
\label{eq:darkState}
\end{equation}

\noindent where $\left|g,1_u,1_l\right>$ is the state with an atom in the ground state $\left|g\right>$ and 1 photon in each of the two cavities. The state described by Eq. \ref{eq:darkState} is similar to the dark state in STIRAP and does not include a component with population in state $\left|i\right>$.

One can excite the state in Eq. \ref{eq:darkState} by initializing the atom in an additional ground state $\left|g_0\right>$ and using an optical drive beam that couples states $\left|g_0\right>$ and $\left|\Psi_0\right>$ (see Appendix C). If the temporal evolution of the drive is slow enough its optical frequency spectrum will only couple to $\left|\Psi_0\right>$ and not to the other eigenstates of the Hamiltonian $H_0$. For $g_l\ll g_u$ and the decay rates $\kappa$ and $\gamma$ small compared to the system's time scale, one can directly go from state $\left|g_0,0_u,0_l\right>$ to $\left|g,1_u,1_l\right>$. We simulate this process by calculating the dynamics of our system using the Lindblad master equation (see Appendix C). In order to study what happens to state $\left|i\right>$ when this is coupled to states $\left|e\right>$ and $\left|g\right>$ by the upper and lower cavities, in all the simulations we fix the cavity detunings to $\Delta=0$. To better illustrate the analogy with STIRAP, we initially consider the scenario where atomic and cavity decays are negligible compared to the atom-cavity coupling rates. The results for a pulse with a duration of 2$\mu s$ and $(g_u, g_l) = (10, 1) \times 2\pi \mathrm{MHz}$ are shown in Fig. \ref{fig:STIRAP}. Here one observes that the atomic population evolves from $\left|g_0\right>$ to $\left|g\right>$ (Fig.~\ref{fig:STIRAP}a) while generating a single photon in each of the two cavity modes (Fig.~\ref{fig:STIRAP}b).
 
\begin{figure}[ht!]
\centering
\includegraphics[width=1\linewidth]{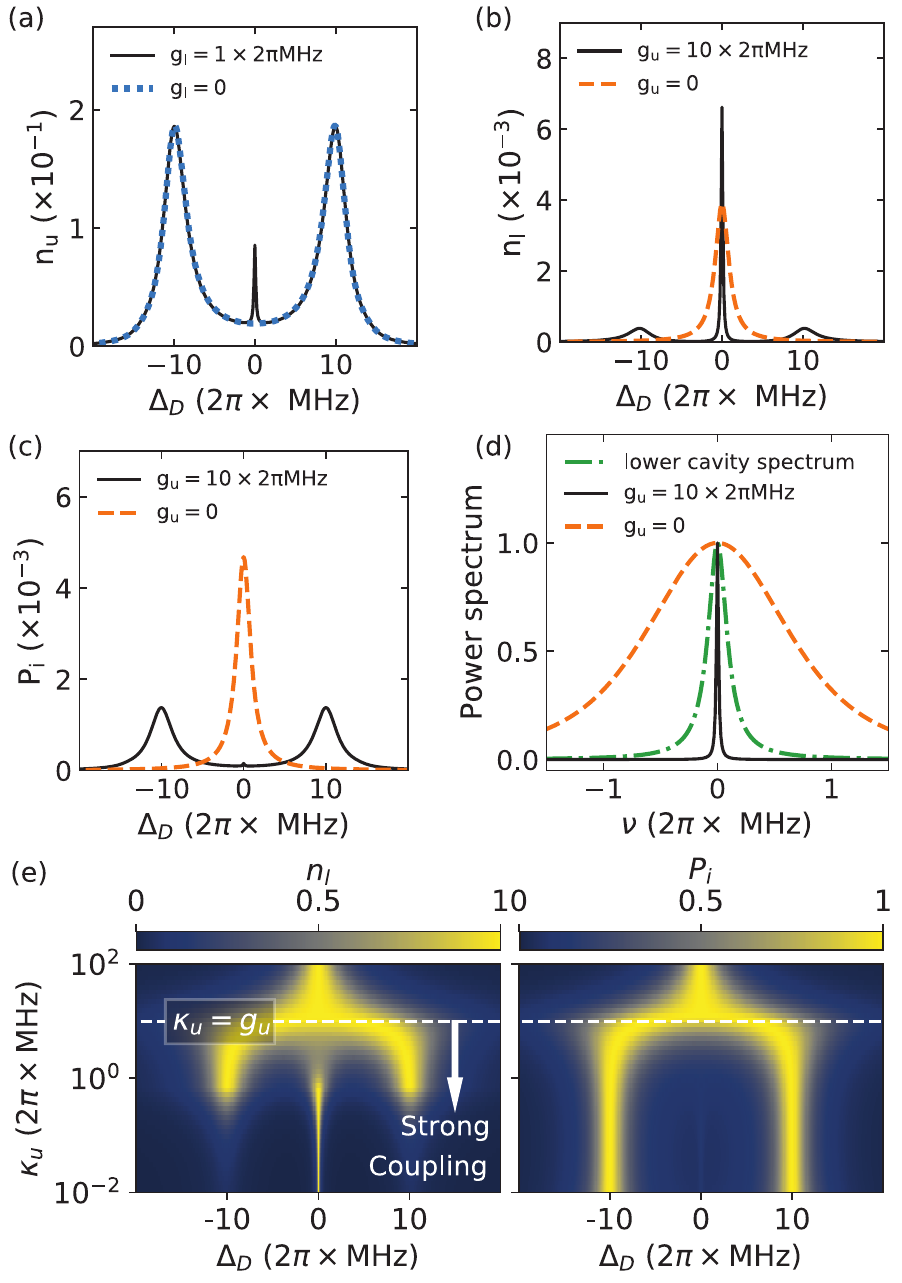}
\caption{Simulation with a driving field coupling the transition $\left|g\right>-\left|e\right>$. (a) Upper cavity photon number, (b) lower cavity photon number and (c) the $\left|i\right>$ state population as a function of the drive field frequency detuning. For comparison the upper (lower) cavity photon numbers are also shown in the situation that the coupling in the other cavity $g_l$ ($g_u$) is null. (d) Spectrum of the lower cavity photon field (black solid line $(g_u, g_l) = (10, 1) \times 2\pi \mathrm{MHz}$, orange dashed line $(g_u, g_l) = (0, 1) \times 2\pi \mathrm{MHz}$) and of the lower empty cavity (green dash-dotted). The other parameters used for this simulation are $(\kappa_u, \kappa_l, \gamma_u, \gamma_l, \Omega_D)$ = (0.01, 0.1, 1, 2, 0.1)$ \times 2\pi \mathrm{MHz}$. (e) Normalized lower cavity photon number $n_l$ and $\left|i\right>$ state population $P_i$ as a function of the upper cavity linewidth and the drive field detuning.}
\label{fig:ssEmission}
\end{figure}

To characterize the mentioned effects in an experiment, we consider some experimentally feasible parameters and simulate the cavity fields, taking into account the above neglected decay rates. We consider a continuous drive on the $\left|g\right>-\left|e\right>$ transition, such that laser excitation and photon emission result in a cycling process between the atomic states (see Appendix C). When we look at the steady-state photon number occupation in the upper cavity as a function of the drive detuning $\Delta_D$ we can see, for $g_l=0$, two photon-generation peaks which correspond to the normal-mode splitting of the $\left|i\right>-\left|e\right>$ transition (see Fig.~\ref{fig:ssEmission}a dashed blue line). However when $g_l\neq 0$ a photon emission peak occurs at $\Delta_D =0$ (Fig.~\ref{fig:ssEmission}a black solid line) due to the presence of the dark state in Eq.~\ref{eq:darkState}. A similar photon emission peak can be observed in the lower cavity for $\Delta_D =0$ when both cavities are coupled to the atom (see Fig.~\ref{fig:ssEmission}b). One can also observe in Fig.~\ref{fig:ssEmission}c that for $\Delta_D=0$ the population in state $\left|i\right>$ is almost zero, much smaller than the population that results when the drive is resonant with one of the normal modes. This observation generalizes a previous finding for one cavity \cite{Villas-Boas2020,Tolazzi2021} instead of our two cavities, namely that a laser-driven cycling scheme involving a cavity vacuum can generate photons without populating the excited state $\left|e\right>$, something we also find in our system. 

One consequence of the small population in state $\left|i\right>$ is that the decay rate of the $\left|i\right>-\left|g\right>$ transition ($\gamma_l$) is not a limiting parameter for the spectrum of the photon. One can then obtain the photon spectrum by calculating the Fourier Transform of the $g^{(1)}$ function of the corresponding field \cite{Khintchine1934}. As illustrated in Fig.~\ref{fig:ssEmission}d, one can observe that for $g_u= 10 \times 2\pi \mathrm{MHz}$ the spectrum of the lower cavity photon is narrower than both the cavity and the transition linewidth, while for $g_u=0$ the spectrum would be significantly broader.

The atom-cavity parameters for the upper transition chosen in this simulation place the system in the strong-coupling regime. As one can see from Fig.~\ref{fig:ssEmission}e, reaching this regime is mandatory to observe the effects just described. In particular, the high-cooperativity regime (i.e. \(C=g^2/(2\kappa\gamma) > 1 \) but not necessarily  $g > (\kappa, \gamma$)) is not sufficient. The more the cavity losses $\kappa_u$ approach the atom-cavity coupling rate $g_u$, the harder it is to resolve the spectral line at $\Delta_D = 0$. Tracing a line at $\kappa_u = g_u$ then allows us to identify two regimes. For $\kappa_u > g_u$ the lower-cavity photon number and the intermediate-state population spectra are nearly the same. In this scenario, the cavity losses erase the coherent transfer from $\left|e\right>$ to $\left|g\right>$. As $\kappa_u$ decreases, the $n_l$ spectrum splits into three peaks, displaying the three eigenstates of the Hamiltonian in eq.~\ref{eq:hamiltonian}. For $g_u \gg \kappa_u$ and $\Delta_D = 0$ photons are generated in the lower cavity while the intermediate state remains unpopulated.

In conclusion, we implemented a photon-pair source using a three-level ladder atom coupled to two optical cavities. We generated pairs of single photons in both cavities with high efficiency, studied their temporal correlation properties and simulated the indistinguishability of both fields. We obtained a probability to emit two single photons (one in each fiber) of $\eta_{\mathrm{pair}}=16(1)\%$. This value is high compared to other state-of-the-art quantum dot photon pair sources, e.g. $\eta_{\mathrm{pair}}=12\%$ \cite{Dousse2010} after the first collecting lens (without any coupling into a single mode fiber), $\eta_{\mathrm{pair}}=0.98\%$ \cite{Wang2019} after coupling the photons in two different fibers, or SPDC sources, e.g. with a photon pair generation probability of $\eta_{\mathrm{pair}}=5\%$ \cite{Zhong2018}. Subsequently we studied theoretically a regime of strong atom-cavity coupling, focusing on a process in which photons are generated without populating the intermediate state. This process is analogous to STIRAP, but mediated by vacuum-cavity fields instead of laser fields. We have studied the properties of the generated photons and proposed a set of parameters in order to observe the effects experimentally. While this work focused on the situation where the two cavities support a single mode of radiation, the scheme could be extended by considering two polarization modes in each cavity. The cavities would then support photonic qubits, and entangled states could be generated in a heralded way \cite{Uphoff2016}.

\begin{acknowledgments}
We thank Carlos Antón-Solanas for useful discussions. This work has been funded by the Deutsche Forschungsgemeinschaft (DFG, German Research Foundation) under Germany’s Excellence Strategy – EXC-2111 – 390814868, and the Bundesministerium für Bildung und Forschung through the Verbund QR.X (16KISQ019).
\end{acknowledgments}

\bigskip

The authors declare no conflicts of interest. Data underlying the results presented in this paper are not publicly available at this time but may be obtained from the authors upon reasonable request.

\bigskip

\appendix

\section{Experimental setup and sequence}

The central part of our experimental setup consists of two crossed optical fiber cavities. The finesse of each cavity is $\mathcal{F}_u=14600$ and $\mathcal{F}_l=15700$, and the cavity lengths are $l_u=162~\mu m$ and $l_l=80~\mu m$, where u denotes the cavity coupled to the upper atomic transition and l the one coupled to the lower atomic transition. The cavity fields have mode waists of
$w_u=6.3~\mu m$ and $w_l=3.5~\mu m,4.8~\mu m$, where the two values for the lower cavity mode correspond to the two axis of the elliptical cavity mode. The two cavities are single sided, meaning that for both cavities the transmission of the outcoupling mirror is $T_{\mathrm{OC}}=340~\mathrm{ppm}$ and the transmission of the high reflective mirror is $T_{\mathrm{HR}}=10~\mathrm{ppm}$. The roundtrip losses for each cavity are $\mathcal{L}_u=80~\mathrm{ppm}$ and $\mathcal{L}_l=50~\mathrm{ppm}$.

We start by preparing a cloud of laser-cooled atoms in a magneto-optical trap around 1~cm above the cavity plane. Upon having created the atomic cloud we let the atoms fall to the intra-cavity region, where we trap individual atoms using a red-detuned standing wave trap at 852~nm (with a waist of 10~$\mu$m) and two blue-detuned intracavity traps at 772~nm and 770~nm. Once the atom is trapped the experimental sequence consists on 20~ms of atom cooling followed by 150~$\mu$s of optical pumping and the atom excitation pulses. The excitation pulse coupling $\left|g_0\right>=\left|5^2S_{1/2},F=1\right>$ to $\left|i\right>=\left|5^2P_{3/2},F=2\right>$ is $\sim$ 1 GHz blue detuned from this transition frequency, whereas the pulse coupling $\left|i\right>$ to $\left|e\right>=\left|5^2D_{5/2},F=3\right>$ is $\sim$ 1 GHz red detuned.  As explained in more detail in the next section, the power of the red-detuned standing wave dipole trap is turned off for 900~ns during the atomic excitation in order to avoid the photoionization of the atom. A feedback loop monitors the photon counts emitted and detected in the lower cavity during the atomic cooling and moves the red-detuned dipole trap standing wave pattern in order to have a single atom in a position of maximal coupling to the cavities. From the rate of detected cooling counts we can determine the number of atoms coupled to the cavities, and in the case that multiple atoms are present the sequence is ceased and the atomic loading is repeated.

The experiment repetition rate ($\sim$ 50 Hz) is intentionally chosen such that the cooling counts acquisition time is sufficiently long to confirm the presence of a single atom at the center of the cavities. A long integration time increases the photon counts signal-to-noise ratio, allowing for more precise positioning of the atom and more accurate post-selection of well positioned atoms.

\section{Atom photoionization}

\begin{figure}
\centering
\includegraphics[width=1\linewidth]{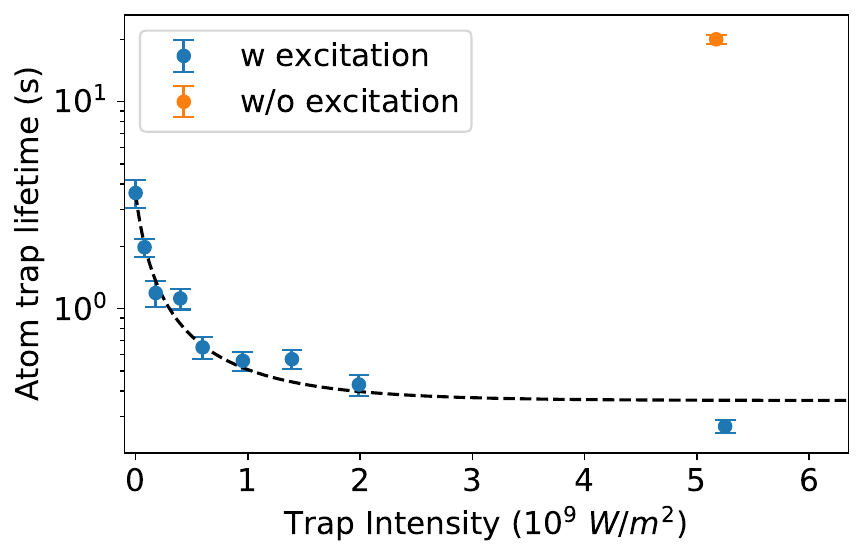}
\caption{Trapped atom lifetime as a function of the trap beam intensity during the atom excitation (blue dots). The black dashed line shows a fit corresponding to a model that considers the photoionization of the atom. The orange dot represents the trap atom lifetime in the case that we do not send the atomic excitation laser pulse.}
\label{fig:photoionization}
\end{figure}

In the experiment described in the main manuscript we excite our $\mathrm{^{87}Rb}$ atom to state $5^2D_{5/2}$. This state is 0.99eV below the ionization energy and therefore the photons in our atomic trap laser with wavelength 852nm have sufficient energy (1.45eV) to induce the photoionization of the atom. We observe this photoionization effect by looking at the atom loss from the trap. Fig.~\ref{fig:photoionization} shows such atom loss effect when sending excitation pulses every 200ms as a function of the trap beam intensity. We can observe how the lifetime of the atom in the trap decreases when we increase the power of the trap during the atom excitation. For this measurement, the trap intensity is kept constant at $5.25 \times 10^9 \mathrm{W/m^2}$ for 200ms and is decreased to the values shown in the plot horizontal axis for 800ns during the atom excitation to $5^2D_{5/2}$. We use the theory described in \cite{Duncan2001} in order to fit the data 

\begin{equation}
P_{\mathrm{PI}}=\eta(1-e^{-\sigma F \lambda/hc})
\label{eq:photoionization}
\end{equation}

where $P_{\mathrm{PI}}$ is the photoionization probability, $\eta$ is the excitation efficiency to the $5D_{5/2}$ level, $\sigma$ is the photoionization cross section, $\lambda$ the light wavelength, and $F=\int^{\infty}_{-\infty}Idt$ is the ionizing fluence related to the light intensity I. From the fit we obtain a photoionization cross section of $\sigma = 17 \pm 6~\mathrm{Mb}$ which is compatible with the photoionization cross section given in the literature for light at 852~nm ($\sigma = 12~\mathrm{Mb}$) \cite{Duncan2001}.

\section{Simulations}

In the main text we define the Hamiltonian that describes our atom-cavity system, which is 

\begin{equation}
H_0 = \Bigl(g_l a_l^{\dagger} \sigma_{gi} + g_u a_u^{\dagger} \sigma_{ie} + \mathrm{H.c.}\Bigr) + \Delta_l a_l^{\dagger} a_l + \Delta_u a_u^{\dagger} a_u
\label{eq:hamiltonian}
\end{equation}

\noindent where $g$ are the atom-photon coupling rates, $a^{\dagger}$ the photonic creation operators, $\sigma^{\dagger}$ the atomic operators, $\Delta$ the atom-cavity detunings and H.c. stands for Hermitian conjugate. The subscripts u and l denote the upper and lower cavity, and the subscripts $g$, $i$ and $e$ denote the atomic levels $\left|g\right>$, $\left|i\right>$ and $\left|e\right>$ as they are described in the main text.

In order to generate photonic fields we need an optical drive that brings energy to the system that is originally in the ground state. In the paper we consider two situations:

\noindent 1) In Fig. 4 (main text) we consider that the atom is initially in an additional ground state $\left|g_0\right>$ and a coupling beam with Rabi frequency $\Omega_D (t)$ which couples states $\left|g_0\right>$ and $\left|e\right>$. This can be described by adding the term in the Hamiltonian 

\begin{equation}
V = \Omega_D (t) (\sigma_{g_0e}+\sigma_{g_0e}^{\dagger})-\Delta_D \left(\sigma_{ee}+a_u^{\dagger}a_u\right)
\label{eq:perturbation1}
\end{equation}
where $\Omega_D(t)$ is the time-dependent driving field Rabi frequency and $\Delta_D$ its detuning to the $\left|g_0\right>-\left|e\right>$ atomic transition.
\newline\newline
\noindent 2) In Fig. 5 (main text) however we don't consider any additional ground state, and we consider a constant driving field with Rabi frequency $\Omega_D$ which couples states $\left|g\right>$ and $\left|e\right>$. In this situation we consider the Hamiltonian in eq.~\ref{eq:hamiltonian} with the additional term
\newline
\begin{equation}
V = \Omega_D (\sigma_{ge}+\sigma_{ge}^{\dagger})-\Delta_D (\sigma_{ee}+a_u^{\dagger}a_u)
\label{eq:perturbation2}
\end{equation}

\noindent In both situations, if the drive field is weak compared to the cavity coupling strengths (i.e. $\Omega_D\ll g_u, g_l$), the term V can be seen as a small perturbation and the system preserves the eigenstates and eigenenergies discussed in the main text.

In order to simulate the dynamics of the system we use the Lindblad master equation 
\begin{multline}
\dot{\rho} = -i \left[H,\rho\right] + \sum_{c = u, l}\kappa_c \left(2 a_c \rho a_c^\dag - a^\dag_c a \rho - \rho a^\dag_c a_c\right) + \\
\gamma_u\Bigl(2\sigma_{ie}\rho\sigma_{ei} - \sigma_{ee}\rho - \rho\sigma_{ee}\Bigr) + \gamma_l\left(2\sigma_{gi}\rho\sigma_{ig} - \sigma_{ii}\rho - \rho\sigma_{ii}\right)
\label{eq:lindblad}
\end{multline}

\noindent Here $\rho$ is the density matrix describing the state of the system, $H$ the Hamiltonian we want to study, $\gamma$ is the atomic population decay rate and $\kappa$ is the cavity field decay rate. The master equation can be solved numerically using the Quantum Toolbox in Python QuTip \cite{Johansson2013}.

\section{Cavity and drive detunings characterization}
\begin{figure}
\centering
\includegraphics[width=1\linewidth]{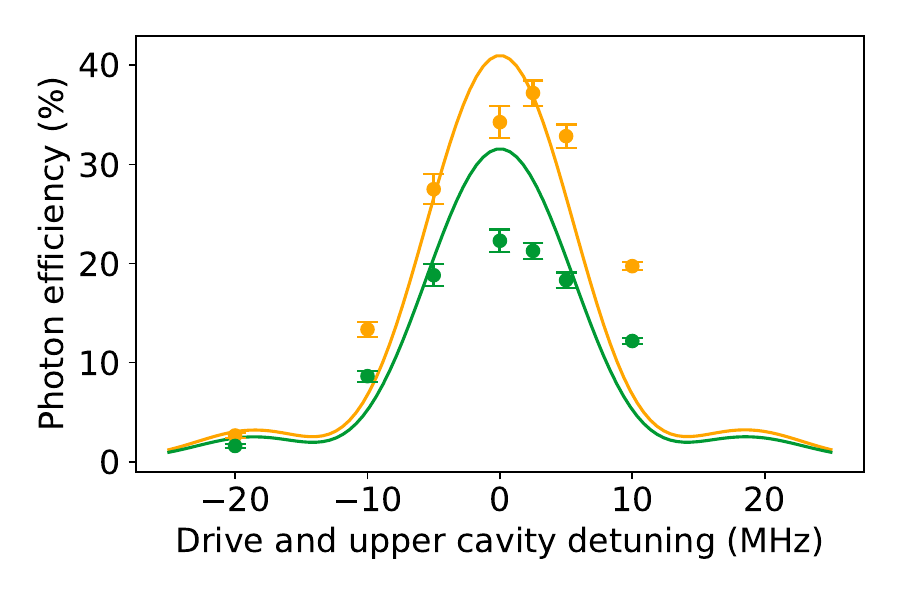}
\caption{ In-fiber photon emission efficiency as a function of the drive beam and upper cavity detuning. The yellow data corresponds to the upper cavity and the green data to the lower cavity. The solid lines show the results of a numerical simulation.}
\label{fig:effDriveDetuning}
\end{figure}

In order to further characterize our photon pair source we study the photon emission dependence as a function of drive and cavity detuning. We start by changing the detuning of the drive field while simultaneous changing the detuning of the upper cavity by the same amount, such that two-photon resonance is preserved. The experimental results are compared to numerical simulations in Fig.~\ref{fig:effDriveDetuning}.

\begin{figure}
\centering
\includegraphics[width=1\linewidth]{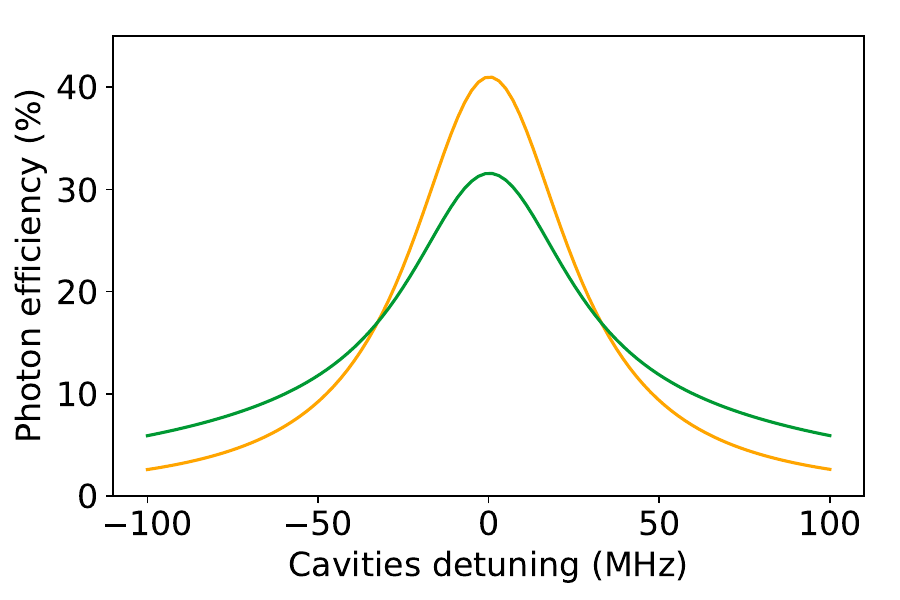}
\caption{ Simulated photon emission efficiency as a function of the upper and lower cavities detuning. The detuning of each cavity has the same absolute value and opposite sign, such that two-photon resonance is preserved. The yellow line corresponds to the upper cavity and the green line to the lower cavity.}
\label{fig:effCavsDetuning}
\end{figure}

We also change the detuning of the two cavities by the same about but with opposite sign, such that the two photon resonance is preserved. In this case we only simulate the described situation, and the results can be seen in Fig.~\ref{fig:effCavsDetuning}. The maximal efficiency happens for zero detuning.

\section{Fiber cavities limitations}
The emission efficiency in the cavity mode $\eta_C$, the cavity-fiber mode matching $\eta_{\mathrm{mm}}$ and the cavity outcoupling efficiency $\eta_{\mathrm{oc}}$ are technically limited by the design of the cavities and cannot be optimized independently from each other.
The cavity-fiber mode matching $\eta_{\mathrm{mm}}$ is optimal if the cavity mode waist is located at the outcoupling mirror and coincides with the fiber mode field diameter. However, this increases the cavity mode waist in the center of the cavity, where the single atoms are trapped, and thus decreases the final cooperativity and emission efficiency of photons into the cavity mode. Further improvements could be achieved by using fiber-integrated mode matching optics that allow for high cavity-fiber mode matching and a small cavity mode waist in the center of the cavity \cite{Gulati2017}.\\
Another limitation is the intracavity losses, primarily due to scattering and absorption at both fiber cavity mirrors (80~ppm and 50~ppm) and transmission through the high-reflective mirror (10~ppm and 10~ppm).\\
To achieve high outcoupling efficiency, the transmission through the outcoupling mirrors (340~ppm and 340~ppm) can be increased at the cost of lower final cooperativity. Therefore, to improve this limitation the intracavity losses must be further reduced by improving the surface qualities of the fiber mirrors during the manufacturing process.


\bibliography{main.bib}

\end{document}